\documentclass{article}

\usepackage{ifthen}
\newboolean{techreport}
\setboolean{techreport}{true}

\newboolean{cameraready}
\setboolean{cameraready}{true}

\usepackage[utf8]{inputenc}
\usepackage[T1]{fontenc}
\usepackage[table]{xcolor}
\usepackage{spconf}
\usepackage[setbb,setcolon]{kmath}
\usepackage{xparse,pgffor,ifthen,xspace}
\usepackage{amsmath,amssymb,amsfonts,amsthm}
\usepackage{mathrsfs,mathtools,bbm}
\usepackage{booktabs,dcolumn,multirow,stmaryrd}
\newcolumntype{d}{D{.}{.}{3.2}}
\newcolumntype{e}{D{.}{.}{1.2}}
\usepackage{pgfplots,forest}
\pgfplotsset{compat=newest}
\usepackage[ruled,linesnumbered]{algorithm2e}
\usepackage{cleveref,csquotes,comment,lipsum,todonotes,soul}

\ifthenelse{\boolean{techreport}}{
	\crefname{section}{Section}{Sections}
	\Crefname{proposition}{Proposition}{Propositions}
	\Crefname{lemma}{Lemma}{Lemmas}
	\Crefname{corollary}{Corollary}{Corollaries}
	\Crefname{appendix}{Appendix}{Appendices}
	\Crefname{figure}{Figure}{Figures}
	\Crefname{equation}{Eq.}{Eqs.}
	\Crefname{algorithm}{Algorithm}{Algorithms}
}{
	\crefname{section}{Sec.}{Sec.}
	\Crefname{proposition}{Prop.}{Prop.}
	\Crefname{lemma}{Lem.}{Lem.}
	\Crefname{corollary}{Cor.}{Cor.}
	\Crefname{appendix}{App.}{App.}
	\Crefname{figure}{Fig.}{Fig.}
	\Crefname{equation}{Eq.}{Eq.}
	\Crefname{algorithm}{Alg.}{Alg.}
}

\usepackage[numbers,sort&compress]{natbib}
\usepackage{glossaries,url,subcaption}

\AfterEndEnvironment{definition}{\noindent\ignorespaces}

\AfterEndEnvironment{remark}{\noindent\ignorespaces}
\newtheorem{proposition}{Proposition}
\AfterEndEnvironment{proposition}{\noindent\ignorespaces}
\newtheorem{lemma}{Lemma}
\AfterEndEnvironment{lemma}{\noindent\ignorespaces}

\AfterEndEnvironment{theorem}{\noindent\ignorespaces}
\newtheorem{corollary}{Corollary}
\AfterEndEnvironment{corollary}{\noindent\ignorespaces}

\AfterEndEnvironment{example}{\noindent\ignorespaces}

\AfterEndEnvironment{problem}{\noindent\ignorespaces}

\foreach \x in {a,...,z}{%
	\expandafter\xdef\csname \x\endcsname{\noexpand\ensuremath{\noexpand\mathbf{\x}}}
}
\foreach \x in {a,...,z}{%
	\expandafter\xdef\csname \x rm\endcsname{\noexpand\ensuremath{\noexpand\mathrm{\x}}}
}

\foreach \x in {A,...,Z}{%
	\expandafter\xdef\csname \x\endcsname{\noexpand\ensuremath{\noexpand\mathbf{\x}}}
}
\foreach \x in {A,...,Z}{%
	\expandafter\xdef\csname \x rm\endcsname{\noexpand\ensuremath{\noexpand\mathrm{\x}}}
}
\foreach \x in {A,...,Z}{%
	\expandafter\xdef\csname \x bb\endcsname{\noexpand\ensuremath{\noexpand\mathbb{\x}}}
}
\foreach \x in {A,...,Z}{%
	\expandafter\xdef\csname \x c\endcsname{\noexpand\ensuremath{\noexpand\mathcal{\x}}}
}

\def\0{{\mathbf 0}}
\def\1{{\mathbf 1}}

\def\ie{\emph{i.e.,}\xspace}
\def\eg{\emph{e.g.,}\xspace}

\ifthenelse{\boolean{cameraready}}{
	\newcommand{\remCE}[1]{}

	\newcommand{\remCH}[1]{}

}{
	\newcommand{\remCE}[1]{{\noindent\color{purple}{\scriptsize[CE: #1]}}}

	\newcommand{\remCH}[1]{{\noindent\color{red}{\scriptsize[CH: #1]}}}

}

\newcommand{\fontset}[1]{\mathcal{#1}}

\newcommand{\fontoptpb}[1]{#1}

\newcommand{\pdim}{n}
\newcommand{\ddim}{m}
\newcommand{\obs}{\y}
\newcommand{\dic}{\A}
\newcommand{\atom}{\a}
\newcommand{\reg}{\lambda}
\newcommand{\bigM}{M}
\newcommand{\pivot}[2]{\boldsymbol{\gamma}_{#1}^{#2}}

\newcommand{\nodeSymb}{\nu}
\newcommand{\nodeSymbIter}[1]{\nodeSymb^{(#1)}}
\newcommand{\setnodesymb}{\fontset{S}}
\newcommand{\setzero}{\setnodesymb_0}
\newcommand{\setone}{\setnodesymb_1}
\newcommand{\setnone}{\bar{\setnodesymb}}

\newcommand{\nodePlusZero}[2]{#1\cup\{\pve_{#2}=0\}}
\newcommand{\nodePlusOne}[2]{#1\cup\{\pve_{#2}\neq0\}}

\newcommand{\subzero}[1]{#1_{\setzero}}

\newcommand{\subnone}[1]{#1_{\setnone}}
\newcommand{\node}[1]{#1^{\nodeSymb}}

\newcommand{\UB}[1]{#1_u}
\newcommand{\LB}[1]{#1_{l}}

\newcommand{\primalletter}{\fontoptpb{P}}
\newcommand{\dualletter}{\fontoptpb{D}}
\newcommand{\mpb}{\primalletter}
\newcommand{\spb}{\primalletter}

\newcommand{\rpb}{\LB{\primalletter}}
\newcommand{\dpb}{\dualletter}

\newcommand{\pobj}{p}
\newcommand{\optobj}{\opt{\pobj}}
\newcommand{\sobj}{\pobj}
\newcommand{\hobj}{\UB{\sobj}}
\newcommand{\robj}{\LB{\sobj}}
\newcommand{\dobj}{d_l}

\newcommand{\dfunc}{\Drm_{l}}
		
\newcommand{\pve}{x}

\newcommand{\dve}{u}
\newcommand{\pv}{\mathbf{\pve}}
\newcommand{\pvrelax}{\mathbf{\pve}_l}

\newcommand{\dv}{\mathbf{\dve}}
\newcommand{\dvopt}{\dv^\star_l}

\newcommand{\idxentrynode}{i}
\newcommand{\idxscreen}{\ell}

\newcommand{\conj}[1]{#1^*}
\newcommand{\card}[1]{|#1|}

\newcommand{\Icvx}{\eta}

\newcommand{\norm}[2]{\|#1\|_#2}
\newcommand{\opt}[1]{#1^{\star}}

\newcommand{\sign}[1]{\mathrm{sign}({#1})}
\newcommand{\relu}[1]{[#1]_+}

\newacronym{bnb}{BnB}{branch-and-bound}
\newacronym{mip}{MIP}{mixed-integer program}
\newcommand{\METHOD}{NODE-SCREENING\xspace}
\newcommand{\Method}{Node-screening\xspace}
\newcommand{\method}{node-screening\xspace}

\usepackage{soul}

\ifthenelse{\boolean{techreport}}{
	\title{\METHOD tests for L0-penalized least-squares problem \\
	With supplementary material
	}
	}{
	\title{\METHOD tests for the L0-penalized least-squares problem}
}

\name{Théo Guyard\(^{\star}\) \qquad Cédric Herzet\(^{\dagger}\) \qquad Clément Elvira\(^{\ddagger}\)}
\address{
	\(^{\star}\) Univ Rennes, INSA Rennes, CNRS, IRMAR-UMR 6625, F-35000 Rennes, France \\ 
	\(^{\dagger}\) INRIA Rennes-Bretagne Atlantique, Campus de Beaulieu, 35000 Rennes, France \\ \(^{\ddagger}\) SCEE/IETR UMR CNRS 6164, CentraleSupélec, 35510 Cesson Sévigné, France \\
	firstname.lastname@\{insa-rennes,inria,centralesupelec\}.fr
	\thanks{
		The research presented in this paper is reproducible.
		Code and data are available at \protect\url{https://gitlab.insa-rennes.fr/Theo.Guyard/bnb-screening}.
	}
}

\begin{document}

\maketitle
\begin{abstract}
	We present a novel screening methodology to safely discard irrelevant nodes within a generic \gls{bnb} algorithm solving the \(\ell_0\)-penalized least-squares problem.
	Our contribution is a set of two simple tests to detect sets of feasible vectors that cannot yield optimal solutions.
	This allows to prune nodes of the \gls{bnb} search tree, thus reducing the overall optimization time.
	One cornerstone of our contribution is a \emph{nesting property} between tests at different nodes that allows to implement them with a low computational cost.
	Our work leverages the concept of safe screening, well known for sparsity-inducing convex problems, and some recent advances in this field for \(\ell_0\)-penalized regression problems.
\end{abstract}


\section{Introduction}

Finding a sparse representation is a fundamental problem in the field of statistics, machine learning and inverse problems.
It consists in decomposing some input vector \(\obs\in\kR^\ddim\) as a linear combination of a few columns (dubbed \emph{atoms}) of a dictionary \(\dic\in\kR^{\ddim\times\pdim}\).
This task can be addressed by solving
\begin{equation}
	\label{prob:prob-base}
	\min_{\pv \in \kR^{\pdim}} \ \tfrac{1}{2}\norm{\obs-\dic\pv}{2}^2 + \reg
	\norm{\pv}{0}
\end{equation}
where
\(\norm{\pv}{0}\) counts the number of nonzero entries in \(\pv\) and \(\reg>0\) is a tuning parameter.
Unfortunately, problem~\eqref{prob:prob-base} has proven to be NP-hard in the general case~\cite[Th.~3]{Chen2012}.
This has led researchers to develop sub-optimal procedures to approximate its solution and address large-scale problems.
Among the most popular, one can mention greedy algorithms~\cite[Sec.~3.2 and Ch.~4]{Foucart2013book}, methodologies based on convex relaxation \cite{selesnick2017sparse,soubies2015continuous} and majorization-minimization algorithms \cite{lanza2017nonconvex}.

Sub-optimal procedures are unfortunately only guaranteed to solve~\eqref{prob:prob-base} under restrictive conditions that are rarely met in practice. 
On the other hand, there has been recently a surge of interest for methods solving~\eqref{prob:prob-base} exactly, see~\cite{Miyashiro2015,bourguignon2015exact,Bertsimas2016,manual1987ibm,mhenni2020sparse} to name a few.
Many approaches leverage the fact that finding
\begin{equation}
	\label{prob:prob} \tag{\(\mpb\)}
	\begin{split}
		\optobj \ =
		\min_{\pv \in \kR^{\pdim}}
		&\tfrac{1}{2} \norm{\obs-\dic\pv}{2}^2 + \reg \norm{\pv}{0}\\
		&\text{s.t.}\,\,\, \kvvbar{\pv}_{\infty} \leq \bigM
	\end{split}
\end{equation}
is equivalent to solve \eqref{prob:prob-base}, provided that \(\bigM\) is chosen large enough.
Interestingly,~\eqref{prob:prob} can be reformulated as a \gls{mip} by introducing binary variables encoding the nullity of the entries of \(\pv\), see \textit{e.g.},~\cite{bourguignon2015exact}.
Besides,~\eqref{prob:prob} has been
shown to be solvable for moderate-size problems by both commercial solvers~\cite{manual1987ibm} and tailored \gls{bnb} algorithms \cite{mhenni2020sparse}.

In a recent paper, Atamtürk and G\'omez extended the notion of safe screening introduced by El Ghaoui \textit{et al.} in~\cite{ghaoui2010safe} from sparsity-promoting convex problems to non-convex \(\ell_0\)-penalized problems.~In particular, they introduced a new methodology that allows to detect (some of) the positions of zero and non-zero entries in the minimizers of a particular~\(\ell_0\)-penalized problem~\cite{atamturk2020safe}.
Their methodology is used as a preprocessing step of any algorithmic procedure and allows
to reduce the problem dimensionality.

In this paper, we make one step forward in the development of numerical methods addressing large-scale \(\ell_0\)-penalized problems by proposing \emph{\method} rules that 
allow to prune \emph{nodes} within the \gls{bnb} search tree.
In contrast to~\cite{atamturk2020safe}, we emphasize the existence of a \emph{nesting property} between screening tests at different nodes.
This enables to (potentially) fix \emph{multiple} entries to either zero or non-zero at any step of the optimization process with a marginal cost.

Our exposition is organized as follows.
\Cref{sec:notations} gathers the main notations used in the paper.
In \Cref{sec:bnb}, we describe a \gls{bnb} algorithm tailored to~\eqref{prob:prob}.
\Cref{sec:screening} presents our new \method tests and explains how to implement them efficiently within the \gls{bnb} process.
In \Cref{sec:results}, we assess the performance of our method on synthetic data.
\ifthenelse{\boolean{techreport}}{
	All the proofs are postponed to Appendix~\ref{sec:appendix proofs}.
}{
	Due to space limitation, the proofs of our results are omitted
	here but can be found in our companion paper~\cite[App.~A]{Guyard2021techreport}.
}

\newcommand{\setnode}{\setfix}

\section{Notations} \label{sec:notations}

We use the following notational conventions throughout the paper.
Boldface uppercase (\textit{e.g.},~\(\dic\)) and lowercase (\textit{e.g.},~\(\pv\)) letters respectively represent matrices and vectors.
\(\0\) denotes the all-zeros vector.
Since the dimension will usually be clear from the context, it is omitted in the notation.
The \(\idxentrynode\)th column of a matrix \(\dic\) is denoted \(\atom_\idxentrynode\).
Similarly, the \(\idxentrynode\)th entry of a vector \(\pv\) is denoted \(\pve_{\idxentrynode}\).
Calligraphic letters (\eg~\(\setnodesymb\)) are used to denote sets and the notation \(\card{\cdot}\) refers to their cardinality.
If \(\setnodesymb\subseteq\{1,\dotsc,\pdim\}\) and \(\pv\in\kR^\pdim\), \(\pv_{\setnodesymb}\) denotes to the restriction of \(\pv\) to its elements indexed by \(\setnodesymb\).
Similarly, \(\dic_{\setnodesymb}\) corresponds to the restriction of \(\dic\in\kR^{\ddim\times\pdim}\) to its columns indexed by \(\setnodesymb\).

\section{Branch-and-bound procedures} \label{sec:bnb}

Branch-and-bound procedures refer to an algorithmic solution to address \glsplural{mip}, among others~\cite{lawler1966branch}.
It consists in implicitly enumerating all feasible solutions and applying pruning rules to discard irrelevant candidates.
When particularized to problem~\eqref{prob:prob}, it can be interpreted as a search tree where a new decision regarding the nullity of an entry of the variable \(\pv\) is taken at each node as illustrated in \Cref{fig:bnb:without screening}.
Formally, we define a node as a triplet \(\nodeSymb=(\setzero,\setone,\setnone)\) where :
\textit{i)} \(\setzero\) and \(\setone\) contain the indices of the entries of $\pv$ which are forced to be zero and non-zero, respectively;
\textit{ii)} \(\setnone\) gathers all the entries of $\pv$ for which no decision has been taken at this stage of the decision tree.

The \gls{bnb} algorithm reads as follows: starting from node \(\nodeSymb=(\emptyset,\emptyset,\{1,\dotsc,\pdim\})\), the method alternates between processing the current node (\emph{bounding} step) and selecting a new node (\emph{branching} step).
The algorithm identifies the global minimum in a finite number of steps with a worst-case complexity equal to an exhaustive search.
In practice, its efficiency depends on both the ability to process nodes quickly and the number of nodes processed.
We review below specific choices for these two steps tailored to problem~\eqref{prob:prob}.

\subsection{Bounding step} \label{subsec:BnB:bounding}

When processing node \(\nodeSymb=(\setzero,\setone,\setnone)\), we prospect if any $\pv$ with zeros on $\setzero$ and non-zero values on $\setone$ can yield a global minimizer of~\eqref{prob:prob}.
To that end, we look for the smallest objective value achieved by feasible candidates with respect to these constraints, \textit{i.e.}, the value of
\begin{equation} \label{prob:prob-node} \tag{\(\node{\spb}\)}
	\begin{split}
		\node{\sobj} \ = \
		\min_{\pv \in \kR^{\pdim}}\ & \tfrac{1}{2} \norm{\obs-\dic\pv}{2}^2 + \reg \norm{\subnone{\pv}}{0} + \reg \card{\setone} \\
		&\mathrm{s.t. } \ \norm{\pv}{\infty} \leq \bigM \ \text{ and }\ \subzero{\pv} = \0.
	\end{split}
\end{equation}
One verifies that \(\node{\sobj} \geq \optobj\) since all minimizers of~\eqref{prob:prob-node} are also feasible for~\eqref{prob:prob}.
Moreover, equality holds whenever the constraints given by \(\setzero\) and \(\setone\) match the configuration of one of the minimizers of \eqref{prob:prob}.

If \(\node{\sobj} > \optobj\), the node \(\nodeSymb\) can be pruned from the tree since no vector $\pv$ verifying
the constraints defined at this node (and therefore at any sub-nodes) can yield an optimal solution.
Unfortunately, the latter pruning rule is of poor practical interest since \(\optobj\) is not available.
Moreover evaluating \(\node{\sobj}\) when \(\setnone\neq\emptyset\) is also  a combinatorial problem.
To circumvent these problems, a relaxed version of the rule is devised: let \(\node{\robj}\) and \(\UB{\pobj}\) be respectively lower and upper bounds on \(\node{\sobj}\) and \(\optobj\). Then, a sufficient condition to prune node $\nodeSymb$ reads \(\node{\robj}> \UB{\pobj}\).

To obtain \(\node{\robj}\), a standard approach~\cite{dong2015regularization} is to consider the following relaxation of \eqref{prob:prob-node} :
\begin{equation} \label{prob:relax-node} \tag{\(\node{\rpb}\)}
	\begin{split}
		\node{\robj}
		\ = \
		\min_{\pv \in \kR^{\pdim}} \ & \tfrac{1}{2} \norm{\obs-\dic\pv}{2}^2 + \tfrac{\reg}{\bigM} \norm{\subnone{\pv}}{1} + \reg \card{\setone} \\
		&\mathrm{s.t. } \ \norm{\pv}{\infty} \leq \bigM \ \text{ and } \ \subzero{\pv} = \0
	\end{split}
\end{equation}
where the \(\ell_0\)-penalty term has been replaced by an \(\ell_1\)-norm.
We emphasize that \(\node{\robj} \leq \node{\sobj}\) since \(\kinv{\bigM}\norm{\subnone{\pv}}{1}\leq\norm{\subnone{\pv}}{0}\) for all feasible vectors \(\pv\).
Interestingly,~\eqref{prob:relax-node} is a constrained LASSO problem~\cite{gaines2018algorithms} and can be solved (to machine precision) in polynomial time by using one of the many methods suitable for this class of problems~\cite{parikh2014proximal,lee2007efficient}.

The upper bound $\UB{\pobj}$ can be obtained by keeping track of the best known objective value of~\eqref{prob:prob} during the \gls{bnb} process.
More precisely, we construct a feasible solution of~\eqref{prob:prob-node} at each node to obtain a potentially better upper bound \(\node{\hobj}\).
The best known upper bound is then updated as \(\UB{\pobj}\gets\min(\UB{\pobj},\node{\hobj})\).
During the \gls{bnb} procedure, \(\UB{\pobj}\) converges toward \(\optobj\) and relaxations are strengthened as new variables are fixed, allowing to prune nodes more effectively.

\subsection{Branching step} \label{subsec:BnB:branching}

If node \(\nodeSymb=(\setzero,\setone,\setnone)\) has not been pruned during the bounding step, the tree exploration goes on.
An index \(\idxentrynode \in \setnone\) is selected according to some \textit{branching rule} and two direct sub-nodes are created below \(\nodeSymb\) by imposing either ``\(\pve_{\idxentrynode} = 0\)'' or ``\(\pve_{\idxentrynode} \neq 0\)''.
Finally, when all nodes have been explored or pruned, the \gls{bnb} algorithm stops and any candidate yielding the best upper bound \(\UB{\sobj}\) is a minimizer of \eqref{prob:prob}.
\begin{figure*}
	\begin{subfigure}[b]{0.49\textwidth}
		\centering
		\scalebox{0.8}{
			\input{img/bnb}
		}
		\caption{
			\label{fig:bnb:without screening}
		}
	\end{subfigure}
	\hfill
	\begin{subfigure}[b]{0.49\textwidth}
		\centering
		\scalebox{0.8}{

\forestset{
    node/.style = {
        draw, 
        circle,
        thick,
        align = center,
        font = \small,
        top color = white,
        bottom color = blue!25,
    },
    branch label/.style={
        edge label = {
            node[midway,fill=white,font=\small,draw=black,text height=0.5em,scale=0.9]{#1}
        }
    },
    bnb/.style={
        branch label,
        for tree = {
            node,
            s sep'+=14mm,
            l sep'+=2mm,
            edge={->, thick, draw opacity=0.2, text opacity = 0.3},
            opacity = 0.25,
            text opacity = 0.25,
        },
        before typesetting nodes={
            for tree={
                split option={content}{:}{content,branch label},
            },
        },
        where n children=0{
            tikz+={
                \draw [thick,dashed,->,opacity=0.25] ([yshift=0pt, xshift=0pt].south east) -- ([yshift=-5pt, xshift=5pt].south east);
                \draw [thick,dashed,->,opacity=0.25]  ([yshift=0pt, xshift=0pt].south west) -- ([yshift=-5pt, xshift=-5pt].south west);
            }
        }{},
    },
}
\begin{forest}
    bnb,
    [\(\nodeSymbIter{0}\),name=root,opacity=1,text opacity=1
        [\(\nodeSymbIter{1}\):\({\pve_{\idxentrynode_1}=0}\)
            [\(\nodeSymbIter{3}\):\({\pve_{\idxentrynode_2}=0}\),name=S3,opacity=1,text opacity=1] {
                \draw [thick,dashed,->] ([yshift=0pt, xshift=0pt].south east) -- ([yshift=-5pt, xshift=5pt].south east);
                \draw [thick,dashed,->]  ([yshift=0pt, xshift=0pt].south west) -- ([yshift=-5pt, xshift=-5pt].south west);
                \draw[thick,->] (root.west) .. controls (-3.8,0) .. (S3.north);
                \node[draw,thick,top color = white,
                bottom color = orange!25,,font=\scriptsize,align=center] at (-3.3,-0.4) (scr) {\bf \Method \\ Set \({\pve_{\idxentrynode_1}=0}\) \\ Set \({\pve_{\idxentrynode_2}=0}\)};
            }
            [\(\nodeSymbIter{4}\):\({\pve_{\idxentrynode_2}\neq0}\)]
        ]
        [\(\nodeSymbIter{2}\):\({\pve_{\idxentrynode_1}\neq0}\),
            [\(\nodeSymbIter{5}\):\({\pve_{\idxentrynode_2}=0}\)]
            [\(\nodeSymbIter{6}\):\({\pve_{\idxentrynode_2}\neq0}\)]
        ]
    ]
\end{forest}
		}
		\caption{
			\label{fig:bnb:with screening}
		}
	\end{subfigure}
	\caption{
		\small
		\label{fig:bnb}
		First nodes of the \gls{bnb} tree.
		The root node \(\nodeSymbIter{0}\) corresponds to problem~\eqref{prob:prob} and each sub-node corresponds to a sub-problem~\eqref{prob:prob-node} with different fixed variables.
		(a)
		Standard implementation of a \gls{bnb} where all nodes are processed.
		(b)
		Impact of applying \method tests on the \gls{bnb} search tree:
		at node \(\nodeSymbIter{0}\), test~\eqref{eq:one-screening-test} is passed for entries \(\idxentrynode_1\) and \(\idxentrynode_2\);
		one can thus directly switch to the sub-node including constraints ``\(\pve_{\idxentrynode_1}=0\)'' and ``\(\pve_{\idxentrynode_2}=0\)'', namely \(\nodeSymbIter{3}\).
		The shaded nodes need not be explored.
	}
\end{figure*}

\section{\METHOD tests} \label{sec:screening}

In this section, we present our new \method tests.
They aim to identify, with marginal cost, nodes of the search tree that \emph{cannot} yield a global minimum of~\eqref{prob:prob}.

\subsection{Dual properties}

The crux of our procedure is a connection between the Fenchel dual problem of~\eqref{prob:relax-node} at two consecutive nodes.
Prior to expose this result, let us define for all \(\dv \in \kR^{\ddim}\) and \(\idxentrynode \in \{1,\dots,\pdim\}\) three families of \emph{pivot values} as
\begin{equation}
	\begin{array}{rcl}
		\pivot{\idxentrynode}{}(\dv)  & \triangleq &
		\bigM(|\ktranspose{\atom}_\idxentrynode\dv| - \tfrac{\reg}{\bigM})      \\
		\pivot{\idxentrynode}{0}(\dv) & \triangleq &
		\bigM\relu{|\ktranspose{\atom}_\idxentrynode\dv| - \tfrac{\reg}{\bigM}} \\
		\pivot{\idxentrynode}{1}(\dv) & \triangleq &
		\bigM\relu{\tfrac{\reg}{\bigM}-|\ktranspose{\atom}_\idxentrynode\dv|}
	\end{array}
\end{equation}
where \(\relu{z} \triangleq \max(0, z)\) for all scalars \(z\).
We now express the Fenchel dual of~\eqref{prob:relax-node} with respect to these quantities.
\begin{proposition}
	\label{prop:dual relaxed problem}
	The Fenchel dual of \eqref{prob:relax-node} is given by
	\begin{equation}
		\label{prob:dual-node}
		\begin{split}
			\node{\dobj} \ = \ \max_{\dv \in \kR^{\ddim}} \node{\dfunc}(\dv) \triangleq& \tfrac{1}{2} \norm{\obs}{2}^2 - \tfrac{1}{2} \norm{\obs-\dv}{2}^2 \\
			&- \textstyle\sum_{\idxentrynode \in \setnone} \pivot{\idxentrynode}{0}(\dv) - \textstyle\sum_{\idxentrynode \in \setone} \pivot{\idxentrynode}{}(\dv)
		\end{split}
		\tag{$\node{\dpb}_l$}
	\end{equation}
	and strong duality holds for \eqref{prob:relax-node}-\eqref{prob:dual-node}, \textit{i.e.}, \(\node{\robj} = \node{\dobj}\).
	Moreover, if \((\opt{\pvrelax},\dvopt)\) is a couple of primal-dual solutions, one has
	\begin{equation} \label{eq:optimality condition}
		\dvopt = \obs - \dic\opt{\pvrelax}
		.
	\end{equation}
\end{proposition}
Hence, at a given node \(\nodeSymb\), the dual objective of~\eqref{prob:relax-node} is the sum of a term common to all nodes and some well-chosen pivot values.
Interestingly, the dual objective function at two consecutive nodes only differs from one pivot value as shown in the following result:
\begin{corollary} \label{prop:dual-link}
	Let \(\nodeSymb=(\setzero,\setone,\setnone)\) and \(\idxscreen \in \setnone\), then \(\forall\dv \in \kR^{\ddim}\),
	\begin{subequations}
		\begin{align}
			\dfunc^{\nodePlusZero{\nodeSymb}{\idxscreen}}(\dv) & = \node{\dfunc}(\dv) + \pivot{\idxscreen}{0}(\dv)
			\label{eq:dual-link:nodeplus0}                                                                         \\
			\dfunc^{\nodePlusOne{\nodeSymb}{\idxscreen}}(\dv)  & = \node{\dfunc}(\dv) + \pivot{\idxscreen}{1}(\dv)
			\label{eq:dual-link:nodeplus1}
		\end{align}
	\end{subequations}
	where \(\nodePlusZero{\nodeSymb}{\idxscreen}\) and \(\nodePlusOne{\nodeSymb}{\idxscreen}\) denote the two \emph{direct} sub-nodes of \(\nodeSymb\) where index \(\idxscreen\) has been swapped from \(\setnone\) to \(\setzero\) or \(\setone\), respectively.
\end{corollary}

\subsection{\Method tests}

We now expose our proposed screening strategy to identify nodes of the tree that provably cannot yield a global minimizer of~\eqref{prob:prob}.
Let \(\nodeSymb\) be a node and \(\UB{\pobj}\) an upper bound on \(\opt{\pobj}\).
We have by strong duality between \eqref{prob:relax-node}-\eqref{prob:dual-node} that
\begin{equation} \label{eq:obj-ordering}
	\forall\dv\in\kR^\ddim,
	\quad
	\node{\dfunc}(\dv) \leq \node{\dobj} = \node{\robj} \leq \node{\sobj}
	.
\end{equation}
Hence, combining~\eqref{eq:obj-ordering} with \Cref{prop:dual-link} leads to the next result:
\begin{proposition} \label{prop:l0-screening-tests}
	Let \(\idxscreen \in \setnone\) be some index.
	Then, \(\forall\dv \in \kR^{\ddim}\),
	\begin{subequations}
		\begin{alignat}{4}
			\node{\dfunc}(\dv) + \pivot{\idxscreen}{0}(\dv) & > \UB{\pobj} & \quad\implies\quad & \sobj^{\nodePlusZero{\nodeSymb}{\idxscreen}} > \opt{\pobj} \label{eq:zero-screening-test} \\
			\node{\dfunc}(\dv) + \pivot{\idxscreen}{1}(\dv) & > \UB{\pobj} & \quad\implies\quad & \sobj^{\nodePlusOne{\nodeSymb}{\idxscreen}} > \opt{\pobj} \label{eq:one-screening-test}
			.
		\end{alignat}
	\end{subequations}
\end{proposition}
Stated otherwise, \Cref{prop:l0-screening-tests} describes a simple procedure to identify some sub-nodes of \(\nodeSymb\) that cannot yield an optimal solution of \eqref{prob:prob}.
In particular, if both~\eqref{eq:zero-screening-test} and~\eqref{eq:one-screening-test} pass for a given index, no feasible vector with respect to the constraints defined at \(\nodeSymb\) is a minimizer of~\eqref{prob:prob}  and \(\nodeSymb\) can therefore be pruned.

Our proposed procedure differs from the pruning methodology described in \Cref{sec:bnb} in several aspects.
Performing a test only requires the evaluation of one single inner product.
They can therefore be implemented at marginal cost compared to the overall cost of the bounding step.
Very importantly, they also inherit from the following \emph{nesting property}:
\begin{corollary}
	\label{cor:nesting property}
	Let \(\nodeSymb=(\setzero,\setone,\setnone)\) be a node. Then, if test~\eqref{eq:zero-screening-test} or~\eqref{eq:one-screening-test} passes for index \(\idxscreen\in\setnone\), then it also passes for any sub-node \(\nodeSymb'=(\setzero',\setone',\setnone')\) of \(\nodeSymb\) 
	such that \(\idxscreen\in\setnone'\).
\end{corollary}
In particular, \Cref{cor:nesting property} has the following consequence.
Assume that two distinct indexes \(\idxscreen\) and \(\idxscreen'\) pass test~\eqref{eq:zero-screening-test} at node \(\nodeSymb\).
Then index \(\idxscreen'\) will also pass~\eqref{eq:zero-screening-test} at node \(\nodePlusOne{\nodeSymb}{\idxscreen}\).
The same consequence holds for test \eqref{eq:one-screening-test} and node \(\nodePlusZero{\nodeSymb}{\idxscreen}\).
In other words, if several \method tests pass at a given node, one can \emph{simultaneously} prune several sub-nodes of \(\nodeSymb\), as illustrated in \Cref{fig:bnb:with screening}.
This is in contrast with the bounding procedure described in \Cref{sec:bnb} which has to process each sub-node individually before applying any potential pruning operation.

\subsection{Implementation considerations}

Implementing \method tests described by \Cref{prop:l0-screening-tests} requires the knowledge of an upper bound \(\UB{\sobj}\) of \(\optobj\) and a suitable \(\dv\in\kR^\ddim\), \ie as close as possible to the maximizer of~\eqref{prob:dual-node}.
The value of \(\UB{\sobj}\) can be obtained at no additional cost since the standard implementation of \gls{bnb} already requires storing such valid upper bounds on \(\opt{\pobj}\).
To obtain a relevant candidate for \(\dv\), we suggest the following procedure: assuming the method used to solve~\eqref{prob:relax-node} generates a sequence of iterates \(\{\pv^{(t)}\}_{t\in\kN}\) that converges to a global minimizer, we set
\begin{equation}
	\label{eq:dual scaling-ish}
	\kforall t\in\kN,\quad
	\dv^{(t)} = \obs - \dic\pv^{(t)}
	.
\end{equation}
This choice enjoys two desirable properties.
First, the sequence \(\{\dv^{(t)}\}_{t\in\kN}\) converges toward a maximizer of~\eqref{prob:dual-node} as a consequence of the optimality condition~\eqref{eq:optimality condition}.
Second, both \(\dv^{(t)}\) and \(\ktranspose{\dic}\dv^{(t)}\) are already evaluated by most solvers as they correspond to the residual error and the (negative) gradient of the least-squares term of the objective function, respectively.~Thus, the computational cost of evaluating the screening tests at all undecided entries is marginal as compared to the cost needed to address~\eqref{prob:relax-node}.
The latter choice for \(\dv^{(t)}\) suggests performing \method tests \emph{during} the bounding step.
Hence, when some entries passes the test at node $\nodeSymb$, the \gls{bnb} algorithm immediately switches to some sub-node, without solving \eqref{prob:relax-node} at high precision.

\section{Numerical results} \label{sec:results}

\newcommand{\sparsitylevel}{k}
\newcommand{\sparsevector}{\pv^0}
\newcommand{\CPLEX}{\texttt{Direct}}
\newcommand{\BNB}{\texttt{BnB}}
\newcommand{\BNBscr}{\texttt{BnB+scr}}

We finally report simulation results illustrating the relevance of the \method methodology on two different setups.

\subsection{Experimental setups}

For each trial, we generate a new realization of \(\dic \in \kR^{\ddim\times\pdim}\) and \(\obs\in\kR^{\ddim}\) as follows.
In the \emph{Gaussian} setup, we set \((\ddim,\pdim)=(500,1000)\) and the entries of \(\dic\) are i.i.d. realizations of a normal distribution.
In the \emph{T\oe{}plitz} setup, we set \((\ddim,\pdim)=(500,300)\) and the first column of \(\dic\) contains a sampled sinc function 
The other columns are obtained by shifting the entries in a row-wise fashion so that \(\dic\) inherits from a T\oe{}plitz structure. 
In both setups, the columns of $\dic$ are normalized to one.
To obtain \(\obs\), we first sample a \(\sparsitylevel\)-sparse vector $\sparsevector\in\kR^\pdim$ with uniformly distributed non-zero entries.
Each non-zero entry is set to \(s(1 + \kvbar{a})\), where \(s\in\{-1,+1\}\) is a random sign and \(a\) is a realization of the normal distribution.
In our experiment, we chose \(\sparsitylevel \in \{5,7,9\}\).
Finally, the observation is constructed as \(\obs = \dic\sparsevector + \boldsymbol{\epsilon}\) 
where the entries of \(\boldsymbol{\epsilon}\) are i.i.d. realizations of a centered Gaussian distribution with standard deviation \(\sigma = \norm{\dic\sparsevector}{2} / \sqrt{10\ddim}\).
Such a design leads to a SNR of \(10\)dB in average~\cite{johnson2006signal}.
We tune \(\reg\) statistically as in~\cite{soussen2011bernoulli}, \textit{i.e.}, by setting \(\reg = 2\sigma^2\log(\pdim/\sparsitylevel - 1)\).
Finally, we empirically set \(\bigM = 1.5\norm{\ktranspose{\dic}\obs}{\infty}\) as advised in~\cite{mhenni2020sparse}.

We compare three methods that address~\eqref{prob:prob}:
\textit{i)} \CPLEX{}, that uses CPLEX~\cite{manual1987ibm}, a state-of-the-art commercial \gls{mip} solver; 
\textit{ii)} \BNB{}, the algorithm presented in~\cite{mhenni2020sparse}, which is (up to our knowledge) the fastest \gls{bnb} algorithm tailored to \eqref{prob:prob}; 
\textit{iii)} \BNBscr{} which corresponds to \BNB{} enhanced with the \method methodology presented in \Cref{sec:screening}.
Note that both \BNB{} and \BNBscr{} leverage an efficient implementation of the Active-Set algorithm~\cite[Sec.~16.5]{wright1999numerical} to solve~\eqref{prob:relax-node} and the branching rule described in~\cite[Sec.~2.2]{mhenni2020sparse}.

Experiments are run on a MacOS with an i7 CPU, clocked at 2.2 GHz and with 8Go of RAM.
We restrict calculations on a single core to avoid bias due to parallelization capabilities.
The Academic Version 20.1 of CPLEX is used and both \BNB{} and \BNBscr{} are implemented in Julia v1.5~\cite{bezanson2012julia}.
Results are averaged over 100 instances of problem \eqref{prob:prob}.

\subsection{Method comparison}

\Cref{tab:figure} compares the average number of nodes treated (\ie the number of relaxations \eqref{prob:relax-node} solved at machine precision) and the optimization time for the three methods and all simulation setups.
One observes that \BNBscr{} outperforms the two other methods on all scenarii and all figures of merit.
We also note that, compared to \BNB{}, the reduction in the optimization time is more significant than the reduction in the number of processed nodes.
A thorough examination of our results indicates that 
the bounding step is performed all the faster as many variables are set to zero in \eqref{prob:relax-node}. Our node-screening methodology allows to reach quickly nodes where the bounding step is performed with a lower computational cost.

As a final remark, we mention that \CPLEX{} relies on an efficient C++ implementation while \BNB{} and \BNBscr{} are implemented in Julia.
As C++ usually runs faster than Julia, there is still room for improvements in the comparison of \BNB{} and \BNBscr{} with \CPLEX{}.


\setlength{\tabcolsep}{2.7pt}

\begin{table}[!ht]
	\small
	\centering
    \begin{tabular}{cc||r|r|r||r|r|r||r|r|r}
	\toprule
	& & \multicolumn{3}{c||}{\CPLEX{}} & \multicolumn{3}{c||}{\BNB{}} & \multicolumn{3}{c}{\BNBscr{}} \\
	& \(\sparsitylevel\) & \multicolumn{1}{c}{N} & \multicolumn{1}{c}{T} & \multicolumn{1}{c||}{F} & \multicolumn{1}{c}{N} & \multicolumn{1}{c}{T} & \multicolumn{1}{c||}{F} & \multicolumn{1}{c}{N} & \multicolumn{1}{c}{T} & \multicolumn{1}{c}{F} \\ \midrule\midrule
    \rowcolor{gray!25} \cellcolor{white} & \(5\) & 96 & 25.9 & 0 & 70 & 1.5 & 0 & \textbf{56} & \textbf{0.7} & \textbf{0} \\
    & \(7\) & 292 & 60.8 & 0 & 180 & 5.1 & 0 & \textbf{152} & \textbf{3.0} & \textbf{0} \\ 
    \rowcolor{gray!25} \cellcolor{white} \parbox[t]{2mm}{\multirow{-3}{*}{\rotatebox[origin=c]{90}{Gaussian}}} & \(9\) & 781 & 102.6 & 10 & 483 & 15.6 & 0 & \textbf{412} & \textbf{9.8} & \textbf{0} \\ \midrule
	\rowcolor{gray!25} \cellcolor{white} & \(5\) & 1,424 & 10.2 & 0 & 965 & 6.4 & 0 & \textbf{725} & \textbf{4.2} & \textbf{0} \\
    & \(7\) & 17,647 & 106.5 & 0 & 10,461 & 79.3 & 0 & \textbf{7,881} & \textbf{52.2} & \textbf{0} \\ 
    \rowcolor{gray!25} \cellcolor{white} \parbox[t]{2mm}{\multirow{-3}{*}{\rotatebox[origin=c]{90}{T\oe{}plitz}}} & \(9\) & 80,694 & 353.4 & 50 & 47,828 & 346.4 & 48 & \textbf{41,166} & \textbf{267.0} & \textbf{40} \\ \bottomrule                      
    \end{tabular}
	\caption{
		\small
		\label{tab:figure}
		Number of nodes explored (N), optimization time in seconds (T) and number of instances not solved within $10^3$ seconds (F).
	}
\end{table}


\section{Conclusion}

In this paper, we presented a novel \method methodology aiming to accelerate the resolution of the \(\ell_0\)-penalized least-squares problem with a \acrlong{bnb} solver.
Our contribution leverages a nesting property between the node-screening tests at two consecutive nodes. 
Our method leads to significant improvements in terms of number of nodes explored and optimization time on two simulated datasets.

\ifthenelse{\boolean{techreport}}{
	\appendix

\section{Proofs} \label{sec:appendix proofs}

\newcommand{\fenchdualvarel}{v}
\newcommand{\fenchdualvar}{\mathbf{\fenchdualvarel}}

This appendix gathers the proofs of the results derived in \Cref{sec:screening}. We refer to~\cite{mhenni2020sparse} for proofs regarding \Cref{sec:bnb}.

\subsection[{Proof of Proposition~\ref{prop:dual relaxed problem}}]{Proof of \Cref{prop:dual relaxed problem}} \label{subsec:proof dual relaxed problem}

Our proof of \Cref{prop:dual relaxed problem} leverages the Fenchel conjugate of a function \(f\) and denoted \(\conj{f}\).
In particular, we first state the following technical lemma whose proof is postponed to the end of the section 
\begin{lemma} \label{lemma:usual-conjugates}
	Define for all \(\z\in\kR^\ddim\) and \(\pv\in\kR^\pdim\)
    \begin{subequations}
        \begin{align}
            f_1(\z) &\triangleq \tfrac{1}{2}\norm{\obs-\z}{2}^2 + \varepsilon
            \label{eq:lemma:conjugate f_1} \\
            f_2(\pv) &\triangleq 
            \tfrac{\reg}{\bigM}\norm{\pv_{\setnodesymb}}{1} 
            + \Icvx_{\kset{\pv'}{\pv'_{\setnodesymb'}=\0}}(\pv)
            + \Icvx_{\mathcal{B}_\infty(M)}(\pv)
            \label{eq:lemma:conjugate f_2} 
        \end{align}
    \end{subequations}
	Then, for all \(\dv\in\kR^\ddim\) and \(\fenchdualvar\in\kR^\pdim\)
    \begin{subequations}
		\begin{align}
			\conj{f_1}(\dv) &= 
			\tfrac{1}{2}(\norm{\obs+\dv}{2}^2 - \norm{\obs}{2}^2) 
			-  \varepsilon \\
			\conj{f_2}(\fenchdualvar) &= 
			\sum_{\idxentrynode\in\setnodesymb}
			\bigM\relu{\kvbar{\fenchdualvarel_\idxentrynode}-\tfrac{\reg}{\bigM}}
			+
			\bigM\norm{\fenchdualvar_{\setnodesymb''}}{1}
			.
		\end{align}
	\end{subequations}
	In the latter result, \(\varepsilon\) is a scalar, \((\setnodesymb,\setnodesymb', \setnodesymb'')\) is a partition of \(\{1,\dotsc,\pdim\}\), \(\mathcal{B}_\infty(M)\) denotes the \(\ell_\infty\)-ball of \(\kR^\pdim\) centered at \(\0\) with radius \(\bigM\), \(\Icvx_{\mathcal{A}}\) denotes the indicator function of set \(\mathcal{A}\subseteq\kR^\pdim\) defined for all \(\pv\in\kR^\pdim\) by \(\Icvx_{\mathcal{A}}(\pv)=0\) if \(\pv\in\mathcal{A}\) and \(+\infty\) otherwise.
\end{lemma}

We now use \Cref{lemma:usual-conjugates} to prove : \textit{a)} the formulation of the dual problem \eqref{prob:dual-node}, \textit{b)} the strong-duality relation linking \eqref{prob:relax-node}-\eqref{prob:dual-node} and \textit{c)} the relation~\eqref{eq:optimality condition}. In the sequel, we let \(\nodeSymb=(\setzero,\setone,\setnone)\) be a given node. \\

\noindent
\textbf{a) Dual problem~\eqref{prob:dual-node}.}
Using the functions defined in \Cref{lemma:usual-conjugates} with \(\varepsilon=\reg\card{\setone}\), \(\setnodesymb=\setnone\), \(\setnodesymb'=\setzero\) and \(\setnodesymb'' = \setone\), the relaxed problem~\eqref{prob:relax-node} can be rewritten as
\begin{equation}
	\label{eq:proof:rewritting relaxed}
    \min_{\pv\in\kR^{\pdim}} \ 
    f_1(\dic\pv) + f_2(\pv)
    .
\end{equation} 
Then, the Fenchel dual of~\eqref{eq:proof:rewritting relaxed} is given by (see~\cite[Definition~15.19]{Bauschke2017}\footnote{Note that, unlike~\cite[Definition~15.19]{Bauschke2017}, we use the formulation of the dual problem that involves a maximization problem.})
\begin{equation}
    \max_{\dv\in\kR^\ddim} \
    - \conj{f_1}(-\dv)
    - \conj{f_2}(\ktranspose{\dic}\dv)
    .
\end{equation}
One concludes the proof by remarking that for all \(\dv\in\kR^\ddim\),
\begin{align}
	\conj{f_2}(\ktranspose{\dic}\dv) - \varepsilon
	\;=\;&
	\sum_{\idxentrynode\in\setnone} \pivot{\idxentrynode}{0}(\dv) 
	+ \bigM\kvvbar{\ktranspose{\dic}_{\setone}\dv}_1 - \tfrac{\bigM}{\bigM} \lambda\card{\setone}
	\nonumber \\
	\;=\;&
	\sum_{\idxentrynode\in\setnone} \pivot{\idxentrynode}{0}(\dv) 
	+ 
	\sum_{\idxentrynode\in\setone} \pivot{\idxentrynode}{}(\dv) 
	\nonumber.
\end{align} \\

\noindent
\textbf{b) Strong duality.}
Strong duality holds as a consequence of~\cite[Proposition 15.24]{Bauschke2017}.
More particularly, one easily verifies that the condition in item~\textit{vii)} is fulfilled since the functions \(f_1\) and \(f_2\) are proper and continuous. \\

\noindent
\textbf{c) Relation~\eqref{eq:optimality condition}.}
Let \((\node{\pvrelax},\node{\dv})\) be a couple of primal-dual solution of~\eqref{prob:relax-node}-\eqref{prob:dual-node}.
Since strong duality holds, we have by item \textit{ii)} of~\cite[Theorem~19.1]{Bauschke2017} that\footnote{Again, we note that~\eqref{eq:proof:dual link opt condition} involves the opposite of \(\node{\dv}\) since the authors consider the reformulation of the dual as a minimization problem.}
\begin{equation}
	\label{eq:proof:dual link opt condition}
	-\node{\dv}\in\partial f_1(\dic\node{\pvrelax})
\end{equation}
where \(\partial f_1(\dic\node{\pvrelax})\) denotes the sub-differential of \(f_1\) evaluated at \(\dic\node{\pvrelax}\).
One finally obtains~\eqref{prop:dual-link} by noticing that \(f_1\) is differentiable at \(\dic\node{\pvrelax}\) so that the sub-differential reduces to the singleton \(\{\nabla f_1(\dic\node{\pvrelax})\}\).

\vspace*{.5em}

\subsection{Proof of \Cref{lemma:usual-conjugates}.}
The Fenchel conjugate of \(f_1\) evaluated at \(\dv\in\kR^{\ddim}\) is defined as
\begin{equation}
	\label{eq:proof:fenchel dual f_1}
	\conj{f_1}(\dv) = \max_{\z\in\kR^{\ddim}}\
	\ktranspose{\dv}\z - f_1(\z)
	.
\end{equation}
One easily sees that~\eqref{eq:proof:fenchel dual f_1} is an unconstrained concave optimization problem whose objective function is differentiable.
We then obtain~\eqref{eq:lemma:conjugate f_1} by noticing that the maximum is attained at
\(\opt{\z}=\obs + \dv\).

\vspace*{.5em}
Let \(\fenchdualvar\in\kR^\pdim\).
By definition of \(\conj{f}_2\) and using the fact that \(f_2(\pv)=+\infty\) for all \(\pv\in\kR^\pdim\) such that \(\pv_{\setnodesymb'}\neq\0\), We have
\begin{align}
	\conj{f}_2(\fenchdualvar)
	\;=\;&
	\max_{
		\pv\in\kintervcc{-\bigM}{\bigM}^\pdim 
	} \,
	\ktranspose{\fenchdualvar}\pv
	- \tfrac{\lambda}{\bigM} \kvvbar{\pv_{\setnodesymb}}_1
	\nonumber 
	\\
	\;=\;&
	\sum_{\idxentrynode\in\setnodesymb}
	\max_{ 
		\pve_\idxentrynode \in \kintervcc{-\bigM}{\bigM}
	}
	\fenchdualvarel_{\idxentrynode}\pve_{\idxentrynode}
	- \tfrac{\lambda}{\bigM} \kvbar{\pve_{\idxentrynode}}
	\nonumber \\
	&\qquad\quad + 
	\max_{ 
		\pv_{\setnodesymb''} \in
		\kintervcc{-\bigM}{\bigM}^{\card{\setnodesymb''}}
	}
	\ktranspose{\fenchdualvar}_{\setnodesymb''}\pv_{\setnodesymb''}
	\nonumber
	.
\end{align}
where the optimization problem can be split since it is linear and \(\setnodesymb,\setnodesymb''\) have non empty intersection by definition.
Let \(\idxentrynode\in\setnodesymb\) and consider the problem
\begin{equation}
	\label{eq:proof:subproblem 1}
	\opt{\pve_{\idxentrynode}} \in \kargmax_{ 
		\pve_\idxentrynode \in \kintervcc{-\bigM}{\bigM}
	} \
	\fenchdualvarel_{\idxentrynode}\pve_{\idxentrynode}
	- \tfrac{\lambda}{\bigM} \kvbar{\pve_{\idxentrynode}}
	.
\end{equation}
Define \(\opt{\pve}_\idxentrynode\) as
\begin{equation}
	\opt{\pve}_\idxentrynode =
	\begin{cases}
		0 &\text{ if } \kvbar{\fenchdualvarel_{\idxentrynode}} - \tfrac{\lambda}{\bigM} \leq 0 \\
		\sign{\fenchdualvarel_{\idxentrynode}} \bigM &\text{ otherwise.}
	\end{cases}
\end{equation}
We let the reader check that there \emph{always} exists \(g\in\partial\kvbar{\opt{\pve}_\idxentrynode}\) such that
\begin{equation}
	\kforall \pve\in\kintervcc{-\bigM}{+\bigM} \quad
	(\fenchdualvarel_{\idxentrynode} - \tfrac{\reg}{\bigM}g) (\pve - \opt{\pve}_\idxentrynode) \leq 0
\end{equation}
Hence, \(\opt{\pve}_\idxentrynode\) is a maximizer of~\eqref{eq:proof:subproblem 1} as a consequence of~\cite[Proposition B.24, item \textit{(f)}]{Bertsekasbook}.
One finally expresses the maximum as a sum of pivot values by injecting the value of \(\opt{\pve}_\idxentrynode\) in the objective function.

Second, see that
\begin{align}
	\max_{ 
		\pv_{\setnodesymb''}\in\kintervcc{-\bigM}{\bigM}^{\card{\setnodesymb''}}
	}
	\ktranspose{\fenchdualvar}_{\setnodesymb''}\pv_{\setnodesymb''} 
	\nonumber 
	\;=\;& 
	\bigM
	\max_{ 
		\pv_{\setnodesymb''}\in\kintervcc{-1}{1}^{\card{\setnodesymb''}}
	}
	\ktranspose{\fenchdualvar}_{\setnodesymb''}\pv_{\setnodesymb''}
	\nonumber \\
	\;=\;& 
	\bigM\kvvbar{\fenchdualvar_{\setnodesymb''}}_1
	\nonumber
\end{align}
where the last equality holds since one recognizes the Fenchel dual of the indicator function of the unit-ball with respect to the \(\ell_\infty\)-norm~\cite[item~\textit{iv)} in Example 13.3]{Bauschke2017} in the penultimate line.

\subsection[Proof of Corollary~\ref{prop:dual-link}]{Proof of \Cref{prop:dual-link}}
	\label{subsec:proof dual link}

Let \(\nodeSymb=(\setzero,\setone,\setnone)\) be a node and \(\idxentrynode\) be some element of \(\setnone\) such that \(\nodePlusZero{\nodeSymb}{\idxentrynode}\) defines a child node.
Using~\eqref{prob:dual-node}, we have for all \(\dv\in\kR^\ddim\)
\begin{align}
	\dfunc^{\nodePlusZero{\nodeSymb}{\idxentrynode}}(\dv)
	-
	\dfunc^{\nodeSymb}(\dv)
	\;=\;&
	\sum_{\idxentrynode' \in \setnone} \pivot{\idxentrynode'}{0}(\dv)
	-
	\sum_{\idxentrynode' \in \setnone\setminus\{\idxentrynode\}} \pivot{\idxentrynode'}{0}(\dv)
	\nonumber \\
	\;=\;&
	\pivot{\idxentrynode}{0}(\dv).
	\nonumber
\end{align}
Similar calculations lead for all \(\dv\in\kR^\ddim\) to
\begin{align}
	\dfunc^{\nodePlusOne{\nodeSymb}{\idxentrynode}}(\dv)
	-
	\dfunc^{\nodeSymb}(\dv)
	\;=\;&
	\pivot{\idxentrynode}{0}(\dv)
	-
	\pivot{\idxentrynode}{}(\dv)
	\nonumber \\
	\;=\,&
	\pivot{\idxentrynode}{1}(\dv)
	\nonumber
\end{align}
where the last line uses the relation \(\relu{\dve} - \relu{-\dve} = \dve\) that hold for all scalar \(\dve\).

\subsection[Proof of Corollary~\ref{cor:nesting property}]{Proof of \Cref{cor:nesting property}} \label{subsec:proof nesting proporty}

Let \(\nodeSymb=(\setzero,\setone,\setnone)\) be a node.
We first state the following lemma which can easily be proved by induction using~\eqref{eq:dual-link:nodeplus0} and~\eqref{eq:dual-link:nodeplus1}. 
\begin{lemma}
	\label{lemma:connection between distant distant child nodes}
	Let \(\nodeSymb'=(\setzero',\setone',\setnone')\) be a child node of \(\nodeSymb\).
	Then, for all \(\dv\in\kR^\ddim\)
	\begin{equation}
		\dfunc^{\nodeSymb'}(\dv)
		=
		\dfunc^{\nodeSymb}(\dv)
		+ \sum_{\idxentrynode\in\setzero'\setminus\setzero} \pivot{\idxentrynode}{0}(\dv)
		+ \sum_{\idxentrynode\in\setone'\setminus\setzero} \pivot{\idxentrynode}{1}(\dv)
		.
	\end{equation}
\end{lemma}
We now prove \Cref{cor:nesting property}.
Let \(\idxscreen\in\setnone\) be some index such that test~\eqref{eq:zero-screening-test} or~\eqref{eq:one-screening-test} passes at node \(\nodeSymb\).
To easier our exposition, we only prove the result for test~\eqref{eq:zero-screening-test} as the same rationale can be used for test~\eqref{eq:one-screening-test}.

Hence, assume that test~\eqref{eq:zero-screening-test} passes for index \(\idxscreen\) at node \(\nodeSymb\), \ie that
\begin{equation}
	 \node{\dfunc}(\dv) + \pivot{\idxscreen}{1}(\dv) > \UB{\pobj}
	 .
\end{equation}
Let \(\nodeSymb'=(\setzero',\setone',\setnone')\) be a child node of \(\nodeSymb\) such that \(\idxscreen\notin\setnone'\).
Using \Cref{lemma:connection between distant distant child nodes}, we have
\begin{equation}
	\dfunc^{\nodeSymb'}(\dv)
	=
	\dfunc^{\nodeSymb}(\dv)
	+ \sum_{\idxentrynode\in\setzero'\setminus\setzero} \pivot{\idxentrynode}{0}(\dv)
	+ \sum_{\idxentrynode\in\setone'\setminus\setzero} \pivot{\idxentrynode}{1}(\dv)
	.
\end{equation}
Since the pivots values \(\{\pivot{\idxentrynode}{0}(\dv)\}_{\idxentrynode=1}^\pdim\)
\(\{\pivot{\idxentrynode}{1}(\dv)\}_{\idxentrynode=1}^\pdim\) are all nonnegative by construction, we necessarily have \(\dfunc^{\nodeSymb'}(\dv) \geq \dfunc^{\nodeSymb}(\dv)\).
Hence the test~\eqref{eq:zero-screening-test} also passes for index \(\idxscreen\) at node \(\nodeSymb'\).

}{
	\clearpage
}

\bibliographystyle{IEEEbib}
\bibliography{refs}

\begin{thebibliography}{10}

\bibitem{Chen2012}
Xiaojun Chen, Dongdong Ge, Zizhuo Wang, and Yinyu Ye,
\newblock ``Complexity of unconstrained $\ell_2-\ell_p$ minimization,''
\newblock {\em Mathematical Programming}, vol. 143, no. 1-2, pp. 371--383,
  November 2014.

\bibitem{Foucart2013book}
Simon Foucart and Holger Rauhut,
\newblock {\em A Mathematical Introduction to Compressive Sensing},
\newblock Springer New York, 2013.

\bibitem{selesnick2017sparse}
Ivan Selesnick,
\newblock ``Sparse regularization via convex analysis,''
\newblock {\em IEEE Transactions on Signal Processing}, vol. 65, no. 17, pp.
  4481--4494, 2017.

\bibitem{soubies2015continuous}
Emmanuel Soubies, Laure Blanc-F{\'e}raud, and Gilles Aubert,
\newblock ``A continuous exact $\ell_0$ penalty $\text{(CEL0)}$ for least
  squares regularized problem,''
\newblock {\em SIAM Journal on Imaging Sciences}, vol. 8, no. 3, pp.
  1607--1639, 2015.

\bibitem{lanza2017nonconvex}
Alessandro Lanza, Serena Morigi, Ivan Selesnick, and Fiorella Sgallari,
\newblock ``Nonconvex nonsmooth optimization via convex--nonconvex
  majorization--minimization,''
\newblock {\em Numerische Mathematik}, vol. 136, no. 2, pp. 343--381, 2017.

\bibitem{Miyashiro2015}
Ryuhei Miyashiro and Yuichi Takano,
\newblock ``Subset selection by {Mallows'} ${C}_p$: A mixed integer programming
  approach,''
\newblock {\em Expert Systems with Applications}, vol. 42, no. 1, pp. 325--331,
  Jan. 2015.

\bibitem{bourguignon2015exact}
S{\'e}bastien Bourguignon, Jordan Ninin, Herv{\'e} Carfantan, and Marcel
  Mongeau,
\newblock ``Exact sparse approximation problems via mixed-integer programming:
  Formulations and computational performance,''
\newblock {\em IEEE Transactions on Signal Processing}, vol. 64, no. 6, pp.
  1405--1419, 2015.

\bibitem{Bertsimas2016}
Dimitris Bertsimas, Angela King, and Rahul Mazumder,
\newblock ``Best subset selection via a modern optimization lens,''
\newblock {\em The Annals of Statistics}, vol. 44, no. 2, Apr. 2016.

\bibitem{manual1987ibm}
CPLEX~User's Manual,
\newblock ``Ibm ilog cplex optimization studio,''
\newblock {\em Version}, vol. 12, pp. 1987--2018, 1987.

\bibitem{mhenni2020sparse}
Ramzi~Ben Mhenni, S{\'e}bastien Bourguignon, Marcel Mongeau, Jordan Ninin, and
  Herv{\'e} Carfantan,
\newblock ``Sparse branch and bound for exact optimization of \(\ell_0\)-norm
  penalized least squares,''
\newblock in {\em ICASSP}. IEEE, 2020, pp. 5735--5739.

\bibitem{ghaoui2010safe}
Laurent~El Ghaoui, Vivian Viallon, and Tarek Rabbani,
\newblock ``Safe feature elimination for the lasso and sparse supervised
  learning problems,''
\newblock 2010.

\bibitem{atamturk2020safe}
Alper Atamturk and Andr{\'e}s G{\'o}mez,
\newblock ``Safe screening rules for l0-regression from perspective
  relaxations,''
\newblock in {\em International conference on machine learning}. PMLR, 2020,
  pp. 421--430.

\bibitem{lawler1966branch}
Eugene~L Lawler and David~E Wood,
\newblock ``Branch-and-bound methods: A survey,''
\newblock {\em Operations research}, vol. 14, no. 4, pp. 699--719, 1966.

\bibitem{dong2015regularization}
Hongbo Dong, Kun Chen, and Jeff Linderoth,
\newblock ``Regularization vs. relaxation: A conic optimization perspective of
  statistical variable selection,''
\newblock {\em arXiv preprint arXiv:1510.06083}, 2015.

\bibitem{gaines2018algorithms}
Brian~R Gaines, Juhyun Kim, and Hua Zhou,
\newblock ``Algorithms for fitting the constrained lasso,''
\newblock {\em Journal of Computational and Graphical Statistics}, vol. 27, no.
  4, pp. 861--871, 2018.

\bibitem{parikh2014proximal}
Neal Parikh and Stephen Boyd,
\newblock ``Proximal algorithms,''
\newblock {\em Foundations and Trends in optimization}, vol. 1, no. 3, pp.
  127--239, 2014.

\bibitem{lee2007efficient}
Honglak Lee, Alexis Battle, Rajat Raina, and Andrew~Y Ng,
\newblock ``Efficient sparse coding algorithms,''
\newblock in {\em Advances in neural information processing systems}, 2007, pp.
  801--808.

\bibitem{johnson2006signal}
Don~H Johnson,
\newblock ``Signal-to-noise ratio,''
\newblock {\em Scholarpedia}, vol. 1, no. 12, pp. 2088, 2006.

\bibitem{soussen2011bernoulli}
Charles Soussen, J{\'e}r{\^o}me Idier, David Brie, and Junbo Duan,
\newblock ``From {Bernoulli-Gaussian} deconvolution to sparse signal
  restoration,''
\newblock {\em IEEE Transactions on Signal Processing}, vol. 59, no. 10, pp.
  4572--4584, 2011.

\bibitem{wright1999numerical}
Stephen Wright, Jorge Nocedal, et~al.,
\newblock ``Numerical optimization,''
\newblock {\em Springer Science}, vol. 35, no. 67-68, pp. 7, 1999.

\bibitem{bezanson2012julia}
Jeff Bezanson, Stefan Karpinski, Viral~B Shah, and Alan Edelman,
\newblock ``Julia: A fast dynamic language for technical computing,''
\newblock 2012.

\bibitem{Bauschke2017}
Heinz~H. Bauschke and Patrick~L. Combettes,
\newblock {\em Convex Analysis and Monotone Operator Theory in Hilbert Spaces},
\newblock Springer International Publishing, 2017.

\bibitem{Bertsekasbook}
Dimitri~P. Bertsekas,
\newblock {\em Nonlinear programming / Dimitri P. Bertsekas},
\newblock Athena Scientific, Belmont (Mass.), 2nd edition edition, 1999.

\end{thebibliography}

\end{document}